\documentclass[10pt,3p,twocolumn]{elsarticle}
\usepackage{amssymb}
\usepackage{url}
\usepackage{chngcntr}
\usepackage{booktabs}
\usepackage{caption}
\biboptions{super}
\usepackage{color}

\begin{document}

\begin{frontmatter}

\title{\LARGE The Landscape of \textit{NeuroImage}-ing Research}

\author[1]{Jordan D. Dworkin\corref{cor1}}
\cortext[cor1]{Corresponding author:}
\ead{jdwor@pennmedicine.upenn.edu}
\author[1]{Russell T. Shinohara}
\author[2,3,4,5]{Danielle S. Bassett}

\address[1]{Department of Biostatistics, Epidemiology, and Informatics, Perelman School of Medicine, University of Pennsylvania, Philadelphia, PA, USA}
\address[2]{Department of Bioengineering, School of Engineering and Applied Sciences, University of Pennsylvania, Philadelphia, PA, USA}
\address[3]{Department of Physics \& Astronomy, College of Arts and Sciences, University of Pennsylvania, Philadelphia, PA, USA}
\address[4]{Department of Electrical \& Systems Engineering, School of Engineering and Applied Sciences, University of Pennsylvania, Philadelphia, PA, USA}
\address[5]{Department of Neurology, Perelman School of Medicine, University of Pennsylvania, Philadelphia, PA, USA}

\begin{abstract}
As the field of neuroimaging grows, it can be difficult for scientists within the field to gain and maintain a detailed understanding of its ever-changing landscape. While collaboration and citation networks highlight important contributions within the field, the roles of and relations among specific areas of study can remain quite opaque. Here, we apply techniques from network science to map the landscape of neuroimaging research documented in the journal \textit{NeuroImage} over the past decade. We create a network in which nodes represent research topics, and edges give the degree to which these topics tend to be covered in tandem. The network displays small-world architecture, with communities characterized by common imaging modalities and medical applications, and with bridges that integrate these distinct subfields. Using node-level analysis, we quantify the structural roles of individual topics within the neuroimaging landscape, and find high levels of clustering within the structural MRI subfield as well as increasing participation among topics related to psychiatry. The overall prevalence of a topic is unrelated to the prevalence of its neighbors, but the degree to which a topic becomes more or less popular over time is strongly related to changes in the prevalence of its neighbors. Broadly, this work presents a cohesive model for understanding the landscape of neuroimaging research across the field, in broad subfields, and within specific topic areas.
\end{abstract}


\end{frontmatter}

\section{Introduction}
\label{intro}
In many fields of research, scientists develop intuitive knowledge of which topics are popular, which might be on the horizon, and which tend to be studied in tandem. Yet each scientist's view of the research landscape is based on a subsampling of the full space, depending on the nature and extent of their experiences in the field. It is therefore often daunting for those who are new to a field to construct even a superficial picture of the research landscape. Moreover, even for those steeped in a particular research area, it can be challenging to imagine new connections that might be drawn between topics that historically have been thought of as unrelated.

Recently, the emerging field of network science has proven useful for gaining an understanding of the broader space of scientific research. Previous work on collaboration and citation networks has provided insight into authors’ social patterns \citep{newman_clustering,moody_netecol}, important turning points in the literature \citep{chen}, and the large-scale structure of the scientific landscape \citep{newman_small,wallace_citation}. But to gain an understanding of the particular landscape of a subfield, it is useful to consider \textit{topic} networks, which reflect the relationships between scientific ideas. Here, the operationalization of science as a set of interconnected ideas provides a unique opportunity to study how research topics are related within and across sub-disciplines, and how these topics and their relations have grown and changed over time.

As neuroimaging researchers, we sought to apply this technique to literature from our field. In this work, we apply graph theoretical approaches to a network of the 100 most common topics covered in the journal \textit{NeuroImage} over the ten-year span from 2008 to 2017. We discuss the large-scale structure of this network, the communities of research areas that emerge from the topic relationships, the roles of individual topics in shaping the network, and the ways in which these roles have changed over time. In sum, our study offers unique insights into the nature and use of scientific research in contemporary neuroimaging.

\section*{Methods}
\subsection*{Data collection}
For this study, we retrieved keywords and abstracts from 8,547 articles published in \textit{NeuroImage} between 2008 and 2017. We used the keyword sections to create a list of potential topics to be searched for in the abstracts. We chose this technique over latent topic modeling for two reasons: (1) it reflected scientists' explicit opinions as to the words and phrases that constitute relevant scientific topics, and (2) it allowed for the incorporation of multi-word phrases. We manually curated the list of topics to eliminate redundancy, unify variations, and account for abbreviations (i.e., \textit{functional mri}, \textit{fMRI}, \textit{functional magnetic resonance imaging}, and \textit{functional magnetic resonance images} were all denoted as \textit{fmri}). Finally, we algorithmically edited abstracts to reflect these changes.

\subsection*{Network construction}
We calculated the prevalence of each potential topic by finding the proportion of abstracts or keyword sections containing the topic phrase within the timespan of study. We used the 100 most common topics between 2008 and 2017 as nodes to construct the network, as this value represented the approximate number at which the least prevalent words occurred sufficiently often to produce a statistically reliable signal. Edges were weighted by the $\phi$ coefficient for binary association \cite{phi}, representing the degree to which two topics tended to be discussed in the same articles. We applied a threshold of positive significance to increase the interpretability of the inter-topic links, and the network that resulted from them. 

\subsection*{Network structure}
To quantify the structural features of the full network, we sought to investigate the degree to which topics tended to form tightly connected clusters, as well as the overall level of integration of research topics across the network.

Local topic clustering can be quantified using the \textbf{clustering coefficient}, which is defined for a node as the probability that two of its adjacent nodes are connected to one another. The version of the clustering coefficient used here is a measure of transitivity defined as follows \cite{barrat}:

$$c_i^{w}=\frac{1}{s_i(k_i-1)}\sum_{h,j\in N} \frac{(w_{ij}+w_{ih})}{2}a_{ij}a_{ih}a_{hj},$$

\noindent where $N$ is the set of all nodes, and $s_{i}$ is the node's \textbf{strength}, or the sum of all edge weights originating at node $i$. The variable $k_{i}$ is the node's \textbf{degree}, or the number of edges originating at node $i$. Finally, $w_{ij}$ is the edge weight connecting node $i$ and node $j$, and $a_{ij}$ is 1 if $w_{ij}>0$ and 0 otherwise. The overall clustering behavior of the network can be obtained by taking the average clustering coefficient over all nodes \cite{bct}.

Integration across the network can be quantified using the \textbf{characteristic path length} of a network. Path length is defined as the average shortest path length between all node pairs\cite{watts_small}. A version of the path length for a weighted network can be defined as follows:

$$L=\frac{1}{n(n-1)}\sum_{i\neq j} d_{ij},$$

\noindent where $n$ is the number of nodes and $d_{ij}$ is the shortest weighted path length between nodes $i$ and $j$.

Notably, these two measures of clustering coefficient and path length can be combined to obtain the \textbf{small-world propensity} of a network, which represents the degree to which a network shows similar clustering to that of a lattice network, and similar path length to that of a random network \cite{swprop}. This metric is similar to the commonly used small-world index \cite{watts_small}, $\sigma$, but has been shown to be unbiased even in the context of networks with varying densities \cite{swprop}. Both measures broadly represent how well a network can be characterized as having both disparate clusters and strong between-cluster integration. The small-world propensity of a network is defined as follows:

$$\phi = 1 - \sqrt{\frac{\Delta_C^2 + \Delta_L^2}{2}},$$

\noindent where

$$\Delta_C = \frac{C_{lattice}-C_{observed}}{C_{lattice}-C_{random}}$$

\noindent and

$$\Delta_L = \frac{L_{observed}-L_{random}}{L_{lattice}-L_{random}},$$

\noindent with $C$ representing the network clustering coefficient, defined as the average of all node-specific $c_i^w$ values.

\subsection*{Community detection}
To determine how the network clustered into subfields, we performed community detection using an iterative generalized Louvain-like locally greedy algorithm to maximize a common modularity quality function \cite{genlouv}. This technique implements a stochastic optimization of the quality index value \textit{Q}, in which nodes are reassigned to modules one by one until no reassignment can improve \textit{Q}. By iterating this optimization until convergence, one obtains a globally optimal set of community assignments, after accounting for local maxima in the \textit{Q} space.

The \textbf{modularity} value, \textit{Q}, of a network represents the degree of separation between nodes in different groups \cite{modul1,modul2}. Intuitively, it quantifies how well the network can be separated into non-overlapping communities with many, strong within-group connections and few, weak between-group connections. For a weighted network, the modularity can be defined as follows:

$$Q^w=\frac{1}{l^w}\sum_{i,j\in N} \left [ w_{ij} - \frac{s_is_j}{l^w} \right ]\delta_{m_im_j},$$

\noindent where $l^w$ is the sum of all edge weights in the network, and $\delta_{m_im_j}$ is 1 if $i=j$ and 0 otherwise. Importantly, it is well known that the modularity landscape suffers from a near degeneracy of optimal solutions \cite{neardeg}. We addressed this issue by using 100 iterations of the Louvain-like algorithm \cite{genlouv}. From the set of partitions obtained across these many iterations, we build a agreement matrix from which we subsequently extract a consensus partition \cite{consensus}.

\subsection*{Node-level structure}
We examined several additional structural features at the level of network nodes. In addition to the previously mentioned measures of node degree and node strength, we sought to understand the ways in which nodes served as bridges between distant topics, and the degree to which they maintained connections to topics outside of their own cluster.

Bridging behavior can be measured using the \textbf{betweenness centrality}, which is the proportion of all shortest paths within the network that pass through a given node \cite{freeman_centrality_1978}. It is defined as follows:

$$b_i=\frac{1}{(n-1)(n-2)}\sum_{h,j\in N; h\neq j, h\neq i, j\neq i} \frac{\rho_{hj}^{(i)}}{\rho_{hj}},$$

\noindent where $\rho_{hj}$ is the number of shortest weighted paths between node $h$ and node $j$, and $\rho_{hj}^{(i)}$ is the number of shortest weighted paths between node $h$ and node $j$ that pass through node $i$.

The diversity of a topic's connections can be measured by the \textbf{participation coefficient}, which quantifies the degree to which the node's connections are evenly distributed across all modules (or clusters) in a network \citep{partic}. The participation coefficient tends towards 1 as connections bridge modules, and tend towards 0 as connections remain largely intramodular. This metric is defined as follows:

$$P_i^w=1-\sum_{m\in M} \left ( \frac{s_i(m)}{s_i} \right )^2,$$

\noindent where $M$ is the set of modules, and $s_i(m)$ is the sum of edge weights from node $i$ to other nodes within module $m$.

To understand how these measures of a node's topological role within the network might relate to the topic's usage in the literature, we examined the prevalence of each topic. We defined the prevalence of a topic as the proportion of article abstracts and keyword sections in which that topic appeared. To further understand how one topic's popularity in the literature might be related to the popularity of topics to which it was closely related, we defined a new measure, $\xi$, that quantified the prevalence of a topic's neighbors. This measure was defined as:

$$\xi_i=\frac{1}{s_i}\sum_{j \neq i} w_{ij} p_j,$$

\noindent where $p_j$ is the overall prevalence of node $j$, given by the proportion of articles in which topic $j$ appears within the time window covered by the given network.

\begin{figure*}
\centering
\includegraphics[width=1\textwidth]{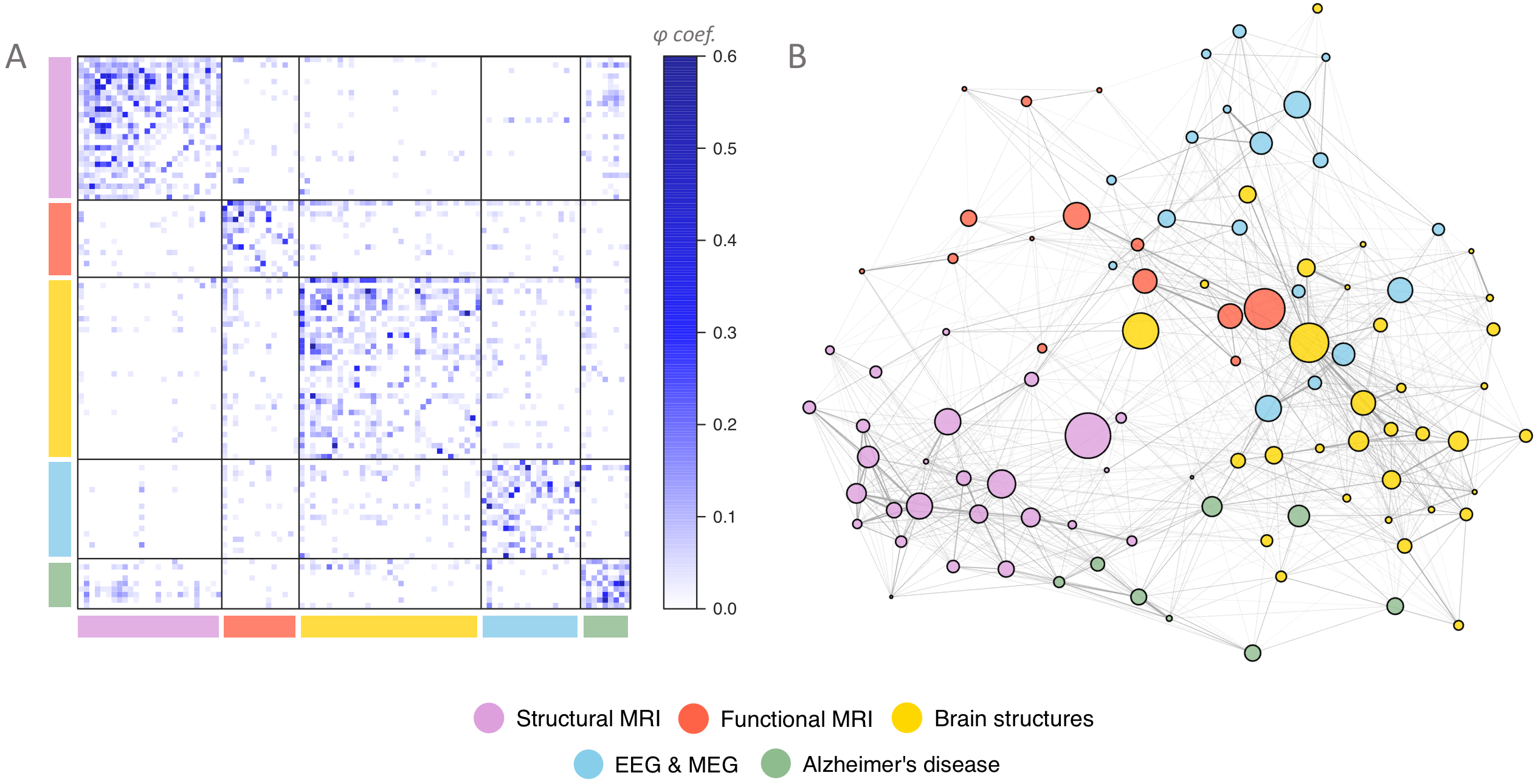}
\caption{\textbf{Architecture of the topic network.} Nodes ($N=100$) reflect research topics and edges ($E=969$) reflect the degree of co-occurrence in abstracts and keyword sections. \textit{(A)} The adjacency matrix sorted by topics' community affiliations, with communities defined by their dominant topic category. \textit{(B)} Visualization of the topic network using the Fruchterman-Reingold layout \cite{frlayout}, where node color represents community affiliation and node size represents prevalence in the literature.}
\label{fig:commsim}
\end{figure*}

\subsection*{Temporal node structure}
To examine changes in topic contributions over time, we created a dynamic network using a sliding window of $\pm$6 months from a central month. Central months ranged from July, 2008 to June, 2017 and data from January, 2008 to December, 2017 were included in the analyses. This dynamic network was comprised of 109 individual networks with overlapping time windows, each with the same 100 topics examined in the full static network.

To examine how topics' structural roles changed over time, we obtained the \textbf{temporal slopes} of each network measure described previously. These slopes were calculated using linear regression on a `month' variable, and were standardized by the magnitude of the measure over the full 10 year time window. The formula is given below, for an example measure, $\theta$:

$$\theta_{j,t}=\alpha+\beta t + \epsilon,$$
$$\Delta^\theta_j=\frac{\hat{\beta}}{\theta_j},$$

\noindent where $\theta_{j,t}$ is the $\theta$ value for node $j$ at the network at time $t$, $\Delta^\theta_j$ represents the slope of measure $\theta$ for node $j$, and $\theta_j$ represents the value of $\theta$ for node $j$ within the context of the full network.

We additionally introduced a measure, $\Omega$, that quantified the degree to which a node is connected to topics that are increasingly used in the literature. It is given by the weighted average of the prevalence slope values of a node's neighbors. It is defined as follows:

$$\Omega_i=\frac{1}{s_i}\sum_{j \neq i} w_{ij} \Delta^p_j,$$

\noindent where $\Delta^p_j$ is the prevalence slope of node $j$, given by the percent change per month in the proportion of articles in which topic $j$ appears, according to the same algorithm given above for calculating slope measures.

\section*{Results}
We collated data from 8,547 articles published in \textit{NeuroImage} between January, 2008 and December, 2017. Using phrases in the keyword sections and their appearance within abstracts, we created a network of research topics. Nodes in the network represented the 100 most common topics over this time span. Edges in the network represented the co-occurrence of topics within articles, quantified by the $\phi$ coefficient of association for binary variables \cite{phi}. We removed negative and non-significant correlations, comprising roughly 63\% of edges; these edges had notably lower magnitude and less variability (range: [-0.23,0.02], interquartile range: 0.02) than the edges that remained (range: [0.02,0.59], interquartile range: 0.06).

\subsection*{Structure of the topic network}
To understand the overall structure of the neuroimaging landscape, we calculated the average node clustering and characteristic path length for the topic network. For comparison, we constructed benchmark distributions from 100 random networks with equivalent size, degree distribution, and strength distribution \cite{rubinov}. We found that the topic network had high levels of clustering ($p<0.01$), with a clustering coefficient 2.24 times the average value found in the null networks. Additionally, the network had a higher characteristic path length than the random networks ($p<0.01$), with a path length 1.31 times the average value of the null networks.

High values of the clustering coefficient and middling values of the characteristic path length can be indicative of small-world architecture in a network. We therefore tested for the presence of small-worldness in the topic network using the small-world propensity score \cite{swprop}. Its value was 0.72, which was significantly higher than would be expected from a random network ($p<0.01$). These results indicate that research topics in neuroimaging have small-world properties, with high local clustering within, and bridges between, distinct subfields. This structure may suggest a pathway for innovation, in which tight clusters advance and refine existing relationships and ideas, and links between clusters facilitate the creation of new connections and future research opportunities \cite{pnas}.

\subsection*{Topic communities}
We next sought to characterize neuroimaging subfields as defined in a data-driven manner from the architecture of the topic network. We performed community detection using a Louvain-like locally greedy algorithm to maximize a common modularity quality function \citep{genlouv}. The modularity value of a network intuitively represents the degree of separation between nodes in different groups \cite{modul1,modul2}. It quantifies how well the network can be separated into non-overlapping communities, with many within-group connections and few between-group connections. Using this technique, we identified five communities, characterized by topics related to structural magnetic resonance imaging (MRI), functional MRI, brain structures, Alzheimer's disease, and electroencephalography (EEG), respectively (Figure \ref{fig:commsim}, \textbf{SI Interactive}).

From the community structure of this network, we were able to identify familiar subfields. Specifically, broad divisions appeared based on imaging modality, with structural MRI, functional MRI, and EEG forming the bases of three separate communities. Additionally, data collection and analysis techniques that are commonly utilized within specific modalities tended to cluster within the modality's community. This was true of the terms \textit{segmentation}, \textit{tractography}, and \textit{voxel-based morphometry} within the structural community. Similarly, it was true of the terms \textit{resting state}, \textit{motion}, and \textit{independent component analysis} within the functional community. And finally, it was also true of the terms \textit{oscillations}, \textit{event-related potential}, and \textit{synchronization} within the EEG community. Intuitively, papers that bridge these communities might combine imaging modalities, or might use the same analysis technique on data acquired in two different imaging modalities.

\begin{figure}
\centering
\includegraphics[width=1\linewidth]{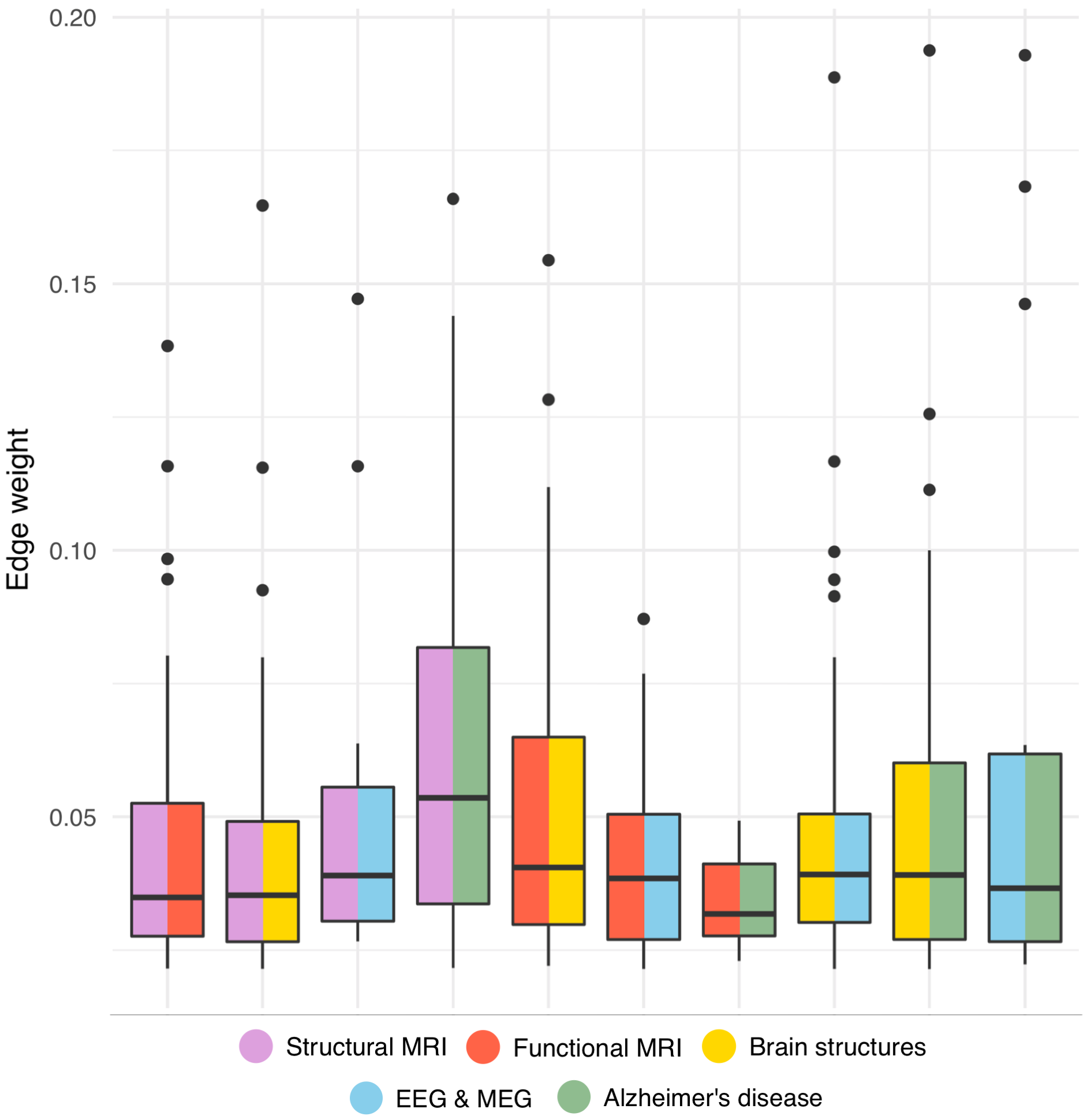}
\caption{\textbf{Strength of cross-community connections.} The distributions of non-zero edge weights for every pair of topic communities. The colors of each boxplot show the pair of communities being represented.}
\label{fig:neighbors}
\end{figure}

While these three communities separated from one another along boundaries of the imaging modality used, the other two communities were more characterized by the content of the neuroscientific question being asked and answered. For example, the largest community, instead of being clustered around a modality, was instead focused on physical structures and associated research topics. This community contained various brain regions, including the \textit{cortex}, \textit{insula}, \textit{cerebellum}, \textit{thalamus}, and \textit{amygdala}, as well as topics that tend to be considered alongside specific regions, such as \textit{emotion}, \textit{depression}, \textit{reward}, and \textit{pain}. Interestingly, the remaining community appeared to be heavily focused on Alzheimer's disease research, including topics like \textit{cognition}, \textit{hippocampus}, \textit{aging}, \textit{atrophy}, and \textit{mild cognitive impairment}. Perhaps unsurprisingly, the closest neighbors of this community were the structural MRI community, and the brain structures community (Figure \ref{fig:neighbors}).

\subsection*{Structural roles of individual topics}
\subsubsection*{Overall structural roles}
To investigate the roles of individual topics in shaping the overall structure of the network, we examined the topics' general connectivity (degree and strength), their role in creating efficient pathways between clusters (betweenness centrality), their level of local clustering (clustering coefficient), and the degree to which they have relationships with topics outside of their own community (participation). See \textbf{SI Interactive} to explore the values of these metrics for each node.

\begin{figure}[!b]
\centering
\includegraphics[width=1\linewidth]{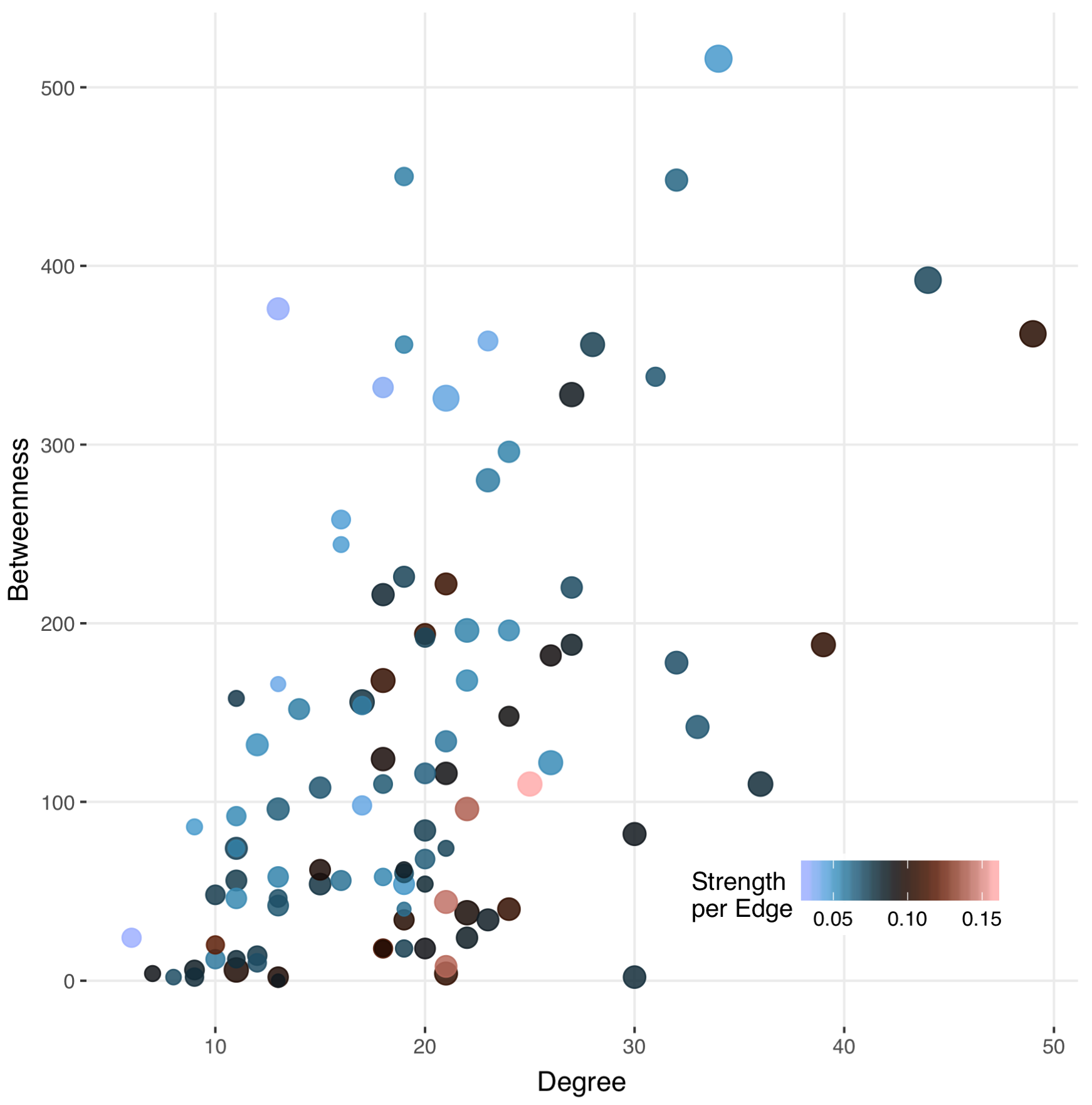}
\caption{\textbf{Relationships between degree, betweenness centrality, and average edge weight.} Points represent individual topics, and are scaled by topic prevalence and colored by the topic's strength per edge.}
\label{fig:between}
\end{figure}

Interestingly, we found that betweenness centrality and degree were significantly related ($\rho=0.49$, $p<0.001$) but betweenness centrality and strength were not ($\rho=0.18$, $p=0.08$). These relationships suggest that topics that serve as bridges connecting disparate communities tended to have many somewhat weak connections. To explore further, we calculated each topic's average strength per edge. This value was significantly negatively related to betweenness centrality ($\rho=-0.37$, $p<0.001$), further supporting the previous finding (Figure \ref{fig:between}). These results are consistent with the presence of broad topics that are covered often and in many different subfields, yielding weak connections to a wide range of other topics. Topics exemplary of this pattern were \textit{brain}, \textit{thalamus}, and \textit{somatosensory cortex}. Topics on the other end of the spectrum, which had high values for average strength per edge, included \textit{white matter}, \textit{diffusion tensor imaging}, and \textit{fractional anisotropy}.

\begin{table*}[ht]
\centering
\small
\begin{tabular}{llllllll}
Rank & Prevalence & Degree & Betweenness & Clustering & Participation \\
\midrule
1 & Brain & Cortex & Brain & DTI & Schizophrenia \\
2 & fMRI & fMRI & Somato. cortex & Multiple sclerosis & Learning \\
3 & Cortex & Prefrontal cortex & Thalamus & Corpus callosum & Hippocampus \\
4 & Human & MRI & fMRI & White matter & Rat \\
5 & MRI & Brain & Motion & Frac. anisotropy & Stroke \\
6 & BOLD & Hippocampus & Cortex & Myelin & Plasticity \\
7 & EEG & Gray matter & Plasticity & Diffusion & Epilepsy \\
8 & Development & Thalamus & Func. connec. & DWI & Medial temp. lobe \\
9 & White matter & Dorso. pf cortex & Depression & Emotion & Memory \\
10 & Memory & Insula & Dorso. pf cortex & Brain development & DM network \\
\bottomrule
\end{tabular}
\caption{\textbf{Rankings of topics' overall structural measures}.}
\label{tab:struc}
\end{table*}

\begin{table*}[ht]
\centering
\small
\begin{tabular}{llllllll}
Rank & $\Delta$ Prevalence & $\Delta$ Degree & $\Delta$ Betweenness & $\Delta$ Clustering & $\Delta$ Participation \\
\midrule
1 & DM network & Dyn. caus. modeling & Synchronization & Plasticity & DTI \\
2 & Resting state & Basal ganglia & Speech & Cerebral cortex & Myelin \\
3 & DWI & Speech & Longitudinal & Somato. cortex & Emotion \\
4 & Func. connec. & Inf. front. gyrus & EEG & CBF & Depression \\
5 & Oscillations & Reading & Reading & Children & Pain \\
6 & Brain development & Reliability & Coherence & DWI & Orbito. cortex \\
7 & Plasticity & Brain development & Corpus callosum & Development & Anxiety \\
8 & TMS & Cerebellum & Auditory cortex & Auditory & Ant. cing. cortex \\
9 & Auditory cortex & Anxiety & ICA & Multiple sclerosis & Amygdala \\
10 & Motion & Alzheimer's disease & Myelin & Segmentation & Hemo. response \\
\bottomrule
\end{tabular}
\caption{\textbf{Rankings of topics' increases in structural measures over time}.}
\label{tab:structemp}
\end{table*}

To understand the trade-off between tight connections within a subfield and broad use connections across the field, we investigated the relationship between topics' clustering coefficient and participation coefficient. Somewhat unsurprisingly, we observed a significant negative relationship between clustering coefficient and participation coefficient ($\rho=-0.66$, $p<0.001$), suggesting that topics with highly related neighbors tend not to have many inter-community connections. Highly clustered topics included \textit{multiple sclerosis}, \textit{corpus collosum}, and \textit{white matter}, while highly participatory topics included \textit{learning}, \textit{hippocampus}, and \textit{schizophrenia} (Table \ref{tab:struc}). The high clustering coefficient throughout the structural MRI community possibly reflects relative isolation of these modalities, methods, and neuroscientific applications within the field.

\subsubsection*{Temporal changes in structural roles}

Though the contributions of individual nodes to the topic network demonstrate the manner in which topics relate to the field as a whole, these metrics only capture a static snapshot of highly dynamic relationships. As such, it was of interest to investigate the dynamic properties of the topic features described in the previous section, in order to better understand how changes in the field may emerge over time. To do so, we created a dynamic network using a $\pm$6-month sliding window that captured the network structure at each central month between July, 2008 and June, 2017. With the $\pm$6-month window, data between January, 2008 and December, 2017 were incorporated into the network.

Dynamic properties of topics were examined by obtaining the slope of each graph metric, and standardizing by the overall metric obtained in the previous section. This calculation yields an estimate of a topic's percent increase or decrease in each measure month-over-month, and can reveal whether topics' roles within the field have been changing over time. The term \textit{connectome} was removed for the following analyses because it was not mentioned in the literature until 2010, making its trajectories strong outliers for almost all measures. Since the metrics that accompanied this topic's entry into the field are still of qualitative interest, the data can be examined in \textbf{SI Interactive}.

As with their absolute magnitudes, we fount that topics' temporal slopes of betweenness centrality and degree were significantly related ($\rho=0.58$, $p<0.001$). However, unlike their absolute magnitudes, slopes of betweenness centrality and strength were also significantly related ($\rho=0.29$, $p<0.01$). This distinction suggests that although topics with the highest betweenness centrality tended to have many weak connections, topics' changes in betweenness centrality were positively related to both the number and strength of their connections. Topics showing increases in betweenness, degree, and strength over time included \textit{speech}, \textit{longitudinal}, and \textit{reading}. Topics that showed decreases in these metrics included \textit{dopamine}, \textit{cerebral blood flow}, and \textit{vision}.

While the trade-off between static clustering coefficient and participation coefficient reflects overall isolation or integration in the past ten years, the changes in these measures over time may better reflect how a topic will relate to the field going forward. Here, the slopes of the clustering coefficient and participation coefficient were negatively related ($\rho=-0.64$, $p<0.001$), suggesting that as topics' become more integrated into a subfield, the strength of their relationships with topics outside of that subfield tends to fade. Topics that have shown increasing clustering over time included \textit{plasticity}, \textit{development}, and \textit{multiple sclerosis}, while topics with increasing participation included \textit{DTI}, \textit{emotion}, and \textit{depression} (Table \ref{tab:structemp}). For this measure, structural MRI topics, while generally high in overall clustering, were found to be moving in a variety of directions. Specifically, topics related to tractography (e.g., \textit{DTI}, \textit{myelin}) showed increasing participation, suggesting increasing integration into the field, while more traditional structural topics (e.g., \textit{segmentation}, \textit{multiple sclerosis}) showed increasing clustering, suggesting further isolation.

\subsection*{Structural foundations of topic prevalence}
\subsubsection*{Overall topic prevalence}
Of particular interest was the relationship between graph measures and topics' popularity in the literature, operationalized by the proportion of articles in which they appeared. In addition to considering the structural features described in the previous section, for this analysis we defined a new variable, $\xi$, which measured the degree to which a topic was a member of a ``rich club" of commonly discussed topics. As opposed to the typical rich club coefficient that measures the edge strengths of each node's neighbors, this metric measured the prevalence of each node's neighbors. We then regressed the log of topic prevalence on betweenness centrality, degree, strength per edge, clustering coefficient, participation coefficient, and $\xi$. We found that degree ($t_{93}=5.29$, $p<0.001$), strength per edge ($t_{93}=4.27$, $p<0.001$), and clustering coefficient ($t_{93}=-3.45$, $p<0.001$) were associated with topic prevalence. These relationships suggest that, accounting for the other variables, topic prevalence tends to be associated with having more, stronger, and more diverse connections. Interestingly, $\xi$ was not associated with prevalence ($t_{93}=0.72$, $p=0.47$), indicating that topics do not tend to be preferably connected to similarly popular topics.

\begin{figure}
\centering
\includegraphics[width=1\linewidth]{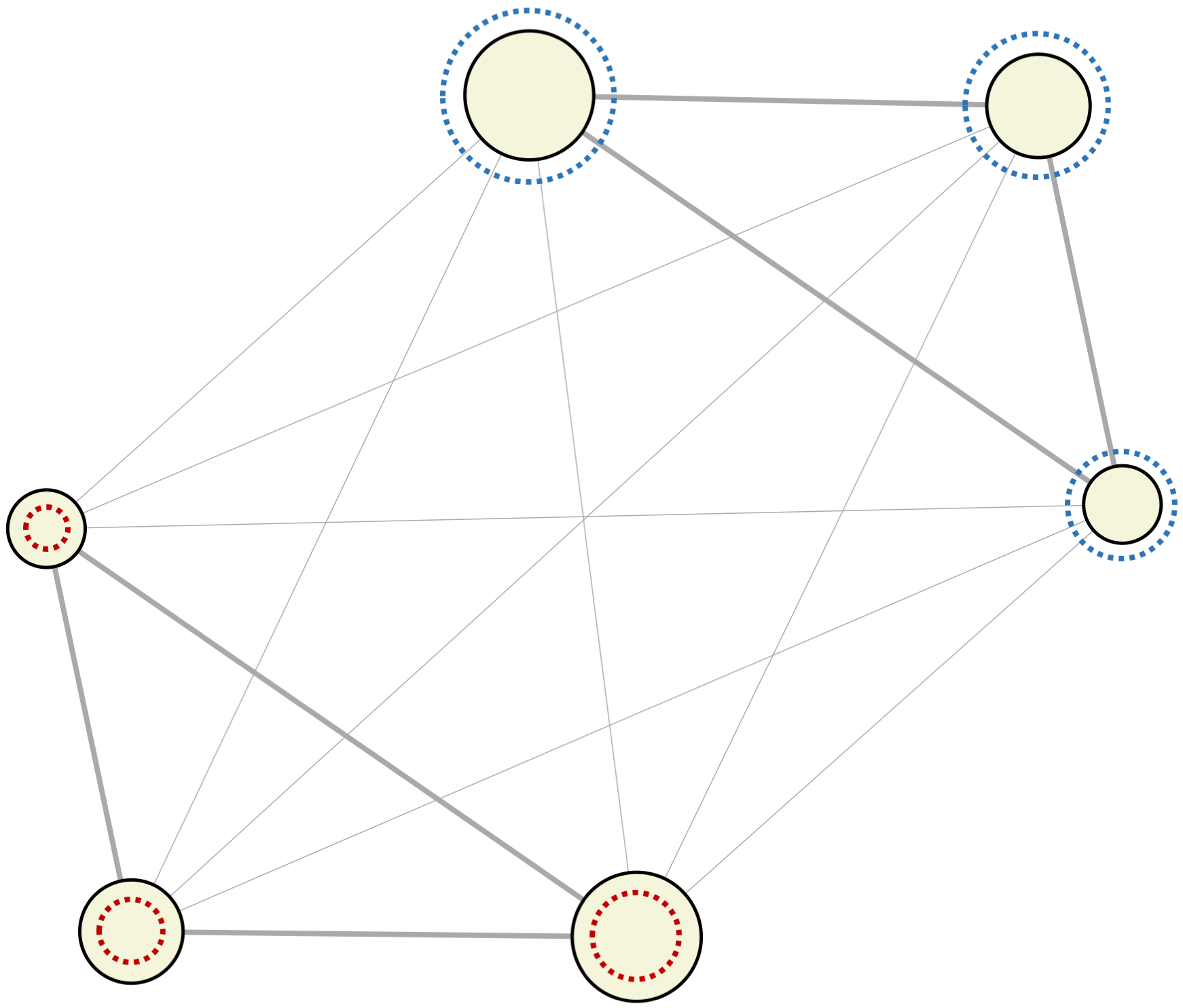}
\caption{\textbf{Illustration of the relationships between topics' prevalence and the prevalence of their neighbors.} Node size represents prevalence, reflecting the finding that topics do not tend to be more strongly connected to topics with similar overall prevalence. Dotted lines represent changes in prevalence over time, reflecting the finding that topics tend to be more strongly connected to topics with similar prevalence change.}
\label{fig:richdiag}
\end{figure}

\subsection*{Temporal changes in topic prevalence}
We next sought to investigate whether changes in topics' structural roles were associated with changes in their prevalence in the literature. Similar to the ``prevalence-rich club" measure just described, for this analysis we defined a new variable, $\Omega$, that measured the degree to which a topic tended to be related to topics with increasing or decreasing prevalence in the literature. We then regressed the slope of topic prevalence on $\Omega$ and the slopes of betweenness centrality, degree, strength per edge, clustering coefficient, and participation coefficient. We found that slopes of degree ($t_{92}=3.30$, $p=0.001$) and participation coefficient ($t_{93}=-2.28$, $p=0.03$) were significantly associated with the slope of topic prevalence. These relationships suggest that topics tend to gain more connections as they become more commonly discussed in the literature, but these connections tend to be limited to the topic's own subfield.

\begin{figure*}
\centering
\includegraphics[width=1\textwidth]{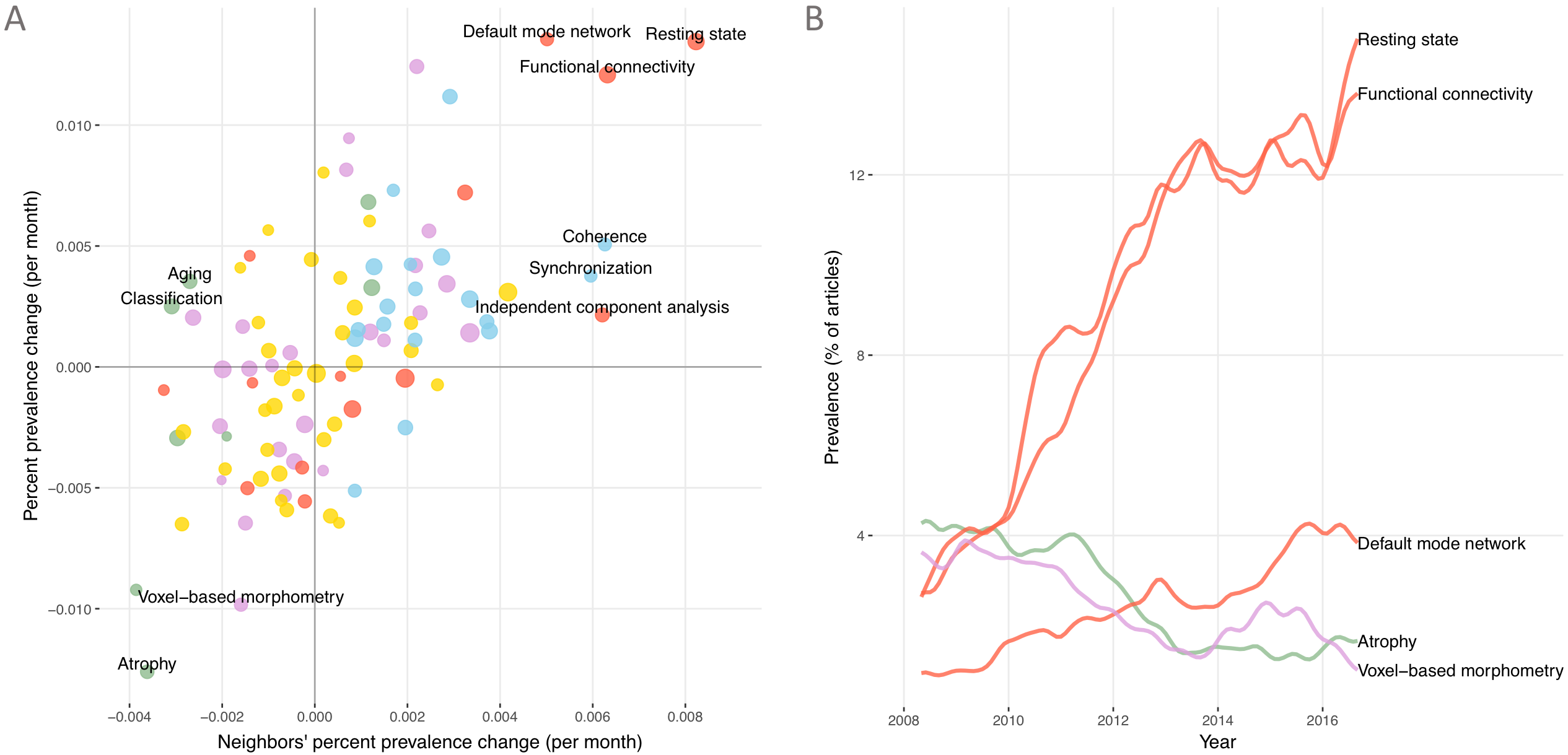}
\caption{\textbf{Changing prevalence of neuroimaging topics.} \textit{(A)} The relationship between topics' changes in prevalence, along the y-axis, and the changes in prevalence of their neighbors, along the x-axis. \textit{(B)} Examples of prevalence trajectories for five exemplary topics with notable increases or decreases in prevalence over the course of the study.}
\label{fig:changeplot}
\end{figure*}

Interestingly, $\Omega$ was strongly associated with changes in topic prevalence ($t_{93}=7.61$, $p<0.001$), indicating that although a topic's overall prevalence is not related to the prevalence of its neighbors, changes in a topic's prevalence are highly related to changes in the prevalence of its neighbors (Figure \ref{fig:richdiag}). This pattern is best exemplified by the trio of topics, \textit{functional connectivity}, \textit{resting state}, and \textit{default mode network}. Certain topics' deviance from the model fit also suggests that -- based on their role in the network -- they might be currently undervalued in the literature. Two such topics, \textit{independent component analysis} and \textit{synchronization}, were predicted to show greater increases in prevalence than was actually observed (Figure \ref{fig:changeplot}).

\section*{Discussion}
Neuroimaging is an exciting, broad, and rapidly changing field. These features make it a rich territory for new ideas and innovative connections between existing topics, but also contribute to a complex and dynamic landscape of research. In this study, we set out to examine the structure of neuroimaging research over the last ten years, and to describe the changing roles of specific areas of study.

\subsection*{Structure of the topic network}
A network of research topics was constructed using ten years of \textit{NeuroImage} articles, and was found to have features uncharacteristic of a random network. Specifically, the network had higher levels of clustering and higher path length than a random network. As a result of the large increase in clustering and moderate increase in path length compared to a random network, the neuroimaging landscape was found to show a high degree of small-worldness. This feature signifies local clustering within distinct subfields, potentially facilitating iteration and advancement within existing topic relationships, and sparse connections between subfields, potentially facilitating innovation and the creation of new connections \cite{newman_main,wagner_gener,pnas}. This small-world structure is consistent with the networks described in studies of co-authorship and citation \cite{newman_small,wallace_citation} at larger scales of scientific research, and interestingly also resembles the structure found in many brain networks \cite{swbrain}.

\subsection*{Topic communities}
While examinations of the network structure of scientific research at a larger scale have revealed interdisciplinary communities of research scientists \cite{comscience} and topics \cite{pnas}, it remains of interest to gain a better understanding of topic communities at the scale of individual fields. At the smaller scale of the neuroimaging topic network, communities appear to largely agree with the subfields that might be expected by a scientist, in that they are divided along the major lines of structural MRI, functional MRI, and EEG/MEG research (Figure \ref{fig:commsim}, \textbf{SI Interactive}). Yet while potentially unsurprising, the strong divisions across imaging modalities suggest that there might be opportunities for further insight and innovation through the joint investigation of several modalities within studies. Indeed, researchers in the field have recently argued that while some specialization is beneficial, experience across modalities is important both professionally and scientifically \cite{headstart}. In some ways, this trade-off reflects a small-scale version of the debate surrounding interdisciplinarity across scientific fields \cite{interdisc,interdisc_assess}, in which some believe better science arises from specialization \cite{davis_grand} and others believe cross-disciplinary work is vital for innovation \cite{nasint}. 

One additional and perhaps surprising finding was the presence of a community that was almost exclusively related to research on Alzheimer's disease. The existence of this community is potentially a reflection of the growing burden of Alzheimer's disease \cite{alzburd}, which has been met with increases in research efforts \cite{alzcent} and neuroimaging resources \cite{adni}. While the prevalence of these topics within neuroimaging appears to indicate a strong response within the field to the pressing need for research in this area, future work could investigate how the characteristics of this sub-field influence the translation of neuroimaging findings to clinical research and practice. This question could potentially be pursued by quantifying the ways in which topic relationships relate to future citation patterns.

\subsection*{Structural roles of individual topics}
In addition to quantifying the overall landscape of neuroimaging, we sought to gain a better understanding of how individual topics contributed to its structure. The degree to which topics bridged gaps between distinct clusters (betweenness centrality) was significantly associated with the number of connections they had to other topics (degree). This pattern of relations is consistent with the presence of ``hub nodes," which have many diverse connections across the network. Yet, interestingly, betweenness centrality was not associated with the strength of these connections, suggesting that hub nodes tended to have relatively weak relationships with the topics that they were connected to. Topics that fit this pattern tended to be very broad, such as \textit{brain} and \textit{cortex}, but some more specific topics also served to connect seemingly distant subfields, like \textit{functional connectivity}, \textit{plasticity}, and \textit{motion} (Table \ref{tab:struc}). 

While patterns in overall betweenness centrality seem to be strongly related to the breadth or specialization of topics, similarly interesting relationships between degree, betweenness centrality, and strength held when looking at their percent change over time within nodes. For these measures, both change in degree and change in strength were positively associated with change in betweenness centrality, suggesting that the associations observed between the overall measures may not be strictly reflective of the inherent nature of the topics, but may instead be tapping into the way topics' structural roles develop and change over time. Additionally, topics that followed this patterns of increasing betweenness centrality and degree over time did not show much overlap with the broad topics that had high overall levels in these measures, and were instead more specific applications like \textit{speech} and \textit{reading} (Table \ref{tab:structemp}).

In terms of the relationships that topics formed with other topics, there seemed to be a trade-off between strong integration in a specific subfield (clustering) and having diverse connections across communities (participation), as evidenced by a significant negative association between the two measures. Topics related to structural MRI tended to be highly clustered, relating very strongly to other topics in that area and rarely forming associations with topics from other subfields. In fact, seven of the ten most clustered topics were directly related to white matter tracts or diffusion imaging. Topics that had diverse connections across subfields were more variable, but tended to be research applications, as opposed to modalities, regions, or methods. Highly participatory topics included \textit{schizophrenia}, \textit{learning}, \textit{stroke}, \textit{epilepsy}, and \textit{memory} (Table \ref{tab:struc}).

The negative association between clustering and participation held for topics' changes in clustering and changes in participation, again suggesting that the overall associations may tap into the manner in which structural roles change alongside topics' changing roles in the literature. Interestingly, although topics related to structural MRI were some of the most highly clustered overall, different topics within this subfield seem to be moving in different directions over time. \textit{Multiple sclerosis} and \textit{segmentation}, for example, showed increasing clustering, while \textit{myelin} and \textit{DTI} showed increasing participation. Applications related to psychiatry also seem to be developing more diverse associations across the neuroimaging landscape, with \textit{emotion}, \textit{depression}, and \textit{anxiety} all showing relative increases in participation (Table \ref{tab:structemp}).

\subsection*{Structural foundations of topic prevalence}
Although the structural roles of topics within the network can provide valuable information about how ideas and subfields interact and change, the aspect that is potentially the most relevant to researchers in the field is the frequency with which topics are discussed in the literature. As such, we sought to understand how a topic's structural features in the neuroimaging network might be associated with its overall prevalence in the literature, and how changes in those features might correspond to changes in its prevalence.

Over the past ten years, we found that a topic's overall prevalence in the literature, defined as the proportion of keyword sections and abstracts in which it was mentioned, was positively associated with its total number of connections and the average strength of its connections, and negatively associated with the degree of local clustering within its neighboring topics. These results suggest that common topics tend to have many relatively strong connections to other topics, and that these neighboring topics tend not to be highly related to each other. Interestingly, highly prevalent topics were not more likely to be connected to other prevalent topics, and therefore little evidence was found for a prevalence-based rich club.

Potentially more interesting than the structural foundations of overall prevalence are the structural foundations of changes in topic prevalence. In other words, we were curious what changes in network features might be associated with changes in how commonly a topic was discussed in the literature. Here, we again found that degree was relevant, with changes in degree showing a positive association with changes in prevalence. Importantly, this finding demonstrates that not only are popular topics also well-connected, but that becoming more well-connected is associated with becoming more popular. Additionally, changes in participation were negatively associated with changes in prevalence, suggesting that as a topic becomes more prevalent, its connections tend to become more localized to the topic's own community. 

Interestingly, the degree to which a topic's neighbors increased or decreased in prevalence was strongly associated with the degree to which that topic itself increased or decreased in prevalence. This pattern of relations is perhaps unsurprising, as one might expect the fortunes of related topics to be linked. Yet the pattern is also potentially quite useful, as it provides a method for determining which topics are out of sync with their neighbors, and therefore might have recently been superseded by other similar topics or might be ripe areas for further research. One example of this phenomenon is \textit{independent component analysis}, which has not seen an increase in prevalence in recent years despite the rise in popularity of many topics with which it is strongly connected (Figure \ref{fig:changeplot}).

\section*{Conclusion}
As science advances, collaboration grows, and the boundaries between research areas blur, it will be increasingly difficult to form and maintain a complete picture of the landscape of any given field. As these changes occur, formal quantitative studies of scientific research offer an opportunity to better synthesize and understand relationships between existing and new domains of inquiry. Here, we used network science to gain insight into the landscape of neuroimaging research in a ten-year span between 2008 and 2017, and revealed previously unknown structural features that emerge from published literature. We found the network to have small-world properties, with communities centered largely around distinct imaging modalities, and bridges between them made up of broadly relevant applications and methods. 

We additionally quantified the structural contributions of individual nodes, finding high clustering among topics related to structural MRI and increasing participation among topics related to psychiatry. Finally, we discovered that the degree to which topics see increasing or decreasing prevalence in the literature is strongly associated with increases or decreases in the prevalence of their neighbors in the network -- a relationship that may help reveal topics that are currently undervalued or recently superseded. Overall, this work sought to characterize the landscape of neuroimaging research at the current moment, and to inform researchers of the structural and literary trends that form the foundations of the field.

\section*{Acknowledgements}
The authors would like to thank Lili Dworkin and Ipek Oguz for advice on data collection and analysis. RTS would like to acknowledge support from the National Institute of Neurological Disorders and Stroke (R01 NS085211 \& R01 NS060910). DSB would like to acknowledge support from the John D. and Catherine T. MacArthur Foundation, the Alfred P. Sloan Foundation, the ISI Foundation, the Paul Allen Foundation, the Army Research Office (DCIST-W911NF-17-2-0181), and the National Science Foundation CAREER (PHY-1554488). The content is solely the responsibility of the authors and does not necessarily represent the official views of any of the funding agencies.

\section*{Supporting Information}
\subsection*{SI Interactive}
A fully interactive version of the network described in this paper can be accessed at https://jdwor.shinyapps.io/NeuroimagingLandscape/ using the password: \textit{NeuroIm10}.

\bibliographystyle{unsrtnat}
\bibliography{bibfile}

\begin{thebibliography}{32}
\providecommand{\natexlab}[1]{#1}
\providecommand{\url}[1]{\texttt{#1}}
\expandafter\ifx\csname urlstyle\endcsname\relax
  \providecommand{\doi}[1]{doi: #1}\else
  \providecommand{\doi}{doi: \begingroup \urlstyle{rm}\Url}\fi

\bibitem[Newman(2001{\natexlab{a}})]{newman_clustering}
M.~E.~J. Newman.
\newblock Clustering and preferential attachment in growing networks.
\newblock \emph{Physical Review E}, 64\penalty0 (2), July 2001{\natexlab{a}}.
\newblock ISSN 1063-651X, 1095-3787.
\newblock \doi{10.1103/PhysRevE.64.025102}.
\newblock URL \url{https://link.aps.org/doi/10.1103/PhysRevE.64.025102}.

\bibitem[Borrett et~al.(2014)Borrett, Moody, and Edelmann]{moody_netecol}
Stuart~R. Borrett, James Moody, and Achim Edelmann.
\newblock The rise of {Network} {Ecology}: {Maps} of the topic diversity and
  scientific collaboration.
\newblock \emph{Ecological Modelling}, 293:\penalty0 111--127, December 2014.
\newblock ISSN 03043800.
\newblock \doi{10.1016/j.ecolmodel.2014.02.019}.
\newblock URL
  \url{http://linkinghub.elsevier.com/retrieve/pii/S0304380014001136}.

\bibitem[Chen(2004)]{chen}
C.~Chen.
\newblock Searching for intellectual turning points: {Progressive} knowledge
  domain visualization.
\newblock \emph{Proceedings of the National Academy of Sciences}, 101\penalty0
  (Supplement 1):\penalty0 5303--5310, April 2004.
\newblock ISSN 0027-8424, 1091-6490.
\newblock \doi{10.1073/pnas.0307513100}.
\newblock URL \url{http://www.pnas.org/cgi/doi/10.1073/pnas.0307513100}.

\bibitem[Newman(2001{\natexlab{b}})]{newman_small}
M.~E.~J. Newman.
\newblock The structure of scientific collaboration networks.
\newblock \emph{Proceedings of the National Academy of Sciences}, 98\penalty0
  (2):\penalty0 404--409, January 2001{\natexlab{b}}.
\newblock ISSN 00278424.
\newblock \doi{10.1073/pnas.021544898}.
\newblock URL \url{\left(}.

\bibitem[Wallace et~al.(2012)Wallace, Larivière, and
  Gingras]{wallace_citation}
Matthew~L. Wallace, Vincent Larivière, and Yves Gingras.
\newblock A {small} {world} of {citations}? {The} {influence} of
  {collaboration} {networks} on {citation} {practices}.
\newblock \emph{PLoS ONE}, 7\penalty0 (3):\penalty0 e33339, March 2012.
\newblock ISSN 1932-6203.
\newblock \doi{10.1371/journal.pone.0033339}.
\newblock URL \url{http://dx.plos.org/10.1371/journal.pone.0033339}.

\bibitem[Davenport and El-Sanhurry(1991)]{phi}
Ernest~C. Davenport and Nader~A. El-Sanhurry.
\newblock Phi/{Phimax}: {Review} and {synthesis}.
\newblock \emph{Educational and Psychological Measurement}, 51\penalty0
  (4):\penalty0 821--828, December 1991.
\newblock ISSN 0013-1644, 1552-3888.
\newblock \doi{10.1177/001316449105100403}.
\newblock URL \url{http://journals.sagepub.com/doi/10.1177/001316449105100403}.

\bibitem[Barrat et~al.(2004)Barrat, Barthelemy, Pastor-Satorras, and
  Vespignani]{barrat}
A.~Barrat, M.~Barthelemy, R.~Pastor-Satorras, and A.~Vespignani.
\newblock The architecture of complex weighted networks.
\newblock \emph{Proceedings of the National Academy of Sciences}, 101\penalty0
  (11):\penalty0 3747--3752, March 2004.
\newblock ISSN 0027-8424, 1091-6490.
\newblock \doi{10.1073/pnas.0400087101}.
\newblock URL \url{http://www.pnas.org/cgi/doi/10.1073/pnas.0400087101}.

\bibitem[Rubinov and Sporns(2010)]{bct}
Mikail Rubinov and Olaf Sporns.
\newblock Complex network measures of brain connectivity: {Uses} and
  interpretations.
\newblock \emph{NeuroImage}, 52\penalty0 (3):\penalty0 1059--1069, September
  2010.
\newblock ISSN 10538119.
\newblock \doi{10.1016/j.neuroimage.2009.10.003}.
\newblock URL
  \url{http://linkinghub.elsevier.com/retrieve/pii/S105381190901074X}.

\bibitem[Watts and Strogatz(1998)]{watts_small}
D.~J. Watts and S.~H. Strogatz.
\newblock Collective dynamics of 'small-world' networks.
\newblock \emph{Nature}, 393\penalty0 (6684):\penalty0 440--442, June 1998.
\newblock ISSN 0028-0836.
\newblock \doi{10.1038/30918}.

\bibitem[Muldoon et~al.(2016)Muldoon, Bridgeford, and Bassett]{swprop}
Sarah~Feldt Muldoon, Eric~W. Bridgeford, and Danielle~S. Bassett.
\newblock Small-{world} {propensity} and {weighted} {brain} {networks}.
\newblock \emph{Scientific Reports}, 6\penalty0 (1), April 2016.
\newblock ISSN 2045-2322.
\newblock \doi{10.1038/srep22057}.
\newblock URL \url{http://www.nature.com/articles/srep22057}.

\bibitem[Jeub et~al.(2011)Jeub, Bazzi, Jutla, and Mucha]{genlouv}
L.G.S. Jeub, M.~Bazzi, I.S. Jutla, and P.J. Mucha.
\newblock A generalized {Louvain} method for community detection implemented in
  {MATLAB}, 2011.
\newblock URL \url{http://netwiki.amath.unc.edu/GenLouvain}.

\bibitem[Newman and Girvan(2004)]{modul1}
M.~E.~J. Newman and M.~Girvan.
\newblock Finding and evaluating community structure in networks.
\newblock \emph{Physical Review E}, 69\penalty0 (2), February 2004.
\newblock ISSN 1539-3755, 1550-2376.
\newblock \doi{10.1103/PhysRevE.69.026113}.
\newblock URL \url{https://link.aps.org/doi/10.1103/PhysRevE.69.026113}.

\bibitem[Newman(2006)]{modul2}
M.~E.~J. Newman.
\newblock Modularity and community structure in networks.
\newblock \emph{Proceedings of the National Academy of Sciences}, 103\penalty0
  (23):\penalty0 8577--8582, June 2006.
\newblock ISSN 0027-8424, 1091-6490.
\newblock \doi{10.1073/pnas.0601602103}.
\newblock URL \url{http://www.pnas.org/cgi/doi/10.1073/pnas.0601602103}.

\bibitem[Good et~al.(2010)Good, de~Montjoye, and Clauset]{neardeg}
Benjamin~H. Good, Yves-Alexandre de~Montjoye, and Aaron Clauset.
\newblock Performance of modularity maximization in practical contexts.
\newblock \emph{Physical Review E}, 81\penalty0 (4), April 2010.
\newblock ISSN 1539-3755, 1550-2376.
\newblock \doi{10.1103/PhysRevE.81.046106}.
\newblock URL \url{https://link.aps.org/doi/10.1103/PhysRevE.81.046106}.

\bibitem[Lancichinetti and Fortunato(2012)]{consensus}
Andrea Lancichinetti and Santo Fortunato.
\newblock Consensus clustering in complex networks.
\newblock \emph{Scientific Reports}, 2\penalty0 (1), December 2012.
\newblock ISSN 2045-2322.
\newblock \doi{10.1038/srep00336}.
\newblock URL \url{http://www.nature.com/articles/srep00336}.

\bibitem[Freeman(1978)]{freeman_centrality_1978}
Linton~C. Freeman.
\newblock Centrality in social networks conceptual clarification.
\newblock \emph{Social Networks}, 1\penalty0 (3):\penalty0 215--239, January
  1978.
\newblock ISSN 03788733.
\newblock \doi{10.1016/0378-8733(78)90021-7}.
\newblock URL
  \url{http://linkinghub.elsevier.com/retrieve/pii/0378873378900217}.

\bibitem[GuimerÃ and Nunes~Amaral(2005)]{partic}
Roger GuimerÃ and LuÃ­s~A. Nunes~Amaral.
\newblock Functional cartography of complex metabolic networks.
\newblock \emph{Nature}, 433\penalty0 (7028):\penalty0 895--900, February 2005.
\newblock ISSN 0028-0836, 1476-4687.
\newblock \doi{10.1038/nature03288}.
\newblock URL \url{http://www.nature.com/articles/nature03288}.

\bibitem[Fruchterman and Reingold(1991)]{frlayout}
Thomas M.~J. Fruchterman and Edward~M. Reingold.
\newblock Graph drawing by force-directed placement.
\newblock \emph{Software: Practice and Experience}, 21\penalty0 (11):\penalty0
  1129--1164, November 1991.
\newblock ISSN 00380644, 1097024X.
\newblock \doi{10.1002/spe.4380211102}.
\newblock URL \url{http://doi.wiley.com/10.1002/spe.4380211102}.

\bibitem[Rubinov and Sporns(2011)]{rubinov}
Mikail Rubinov and Olaf Sporns.
\newblock Weight-conserving characterization of complex functional brain
  networks.
\newblock \emph{NeuroImage}, 56\penalty0 (4):\penalty0 2068--2079, June 2011.
\newblock ISSN 10538119.
\newblock \doi{10.1016/j.neuroimage.2011.03.069}.
\newblock URL
  \url{http://linkinghub.elsevier.com/retrieve/pii/S105381191100348X}.

\bibitem[{Dworkin} et~al.(2018){Dworkin}, {Shinohara}, and {Bassett}]{pnas}
J.~D. {Dworkin}, R.~T. {Shinohara}, and D.~S. {Bassett}.
\newblock {The emergent integrated network structure of scientific research}.
\newblock \emph{ArXiv e-prints}, April 2018.

\bibitem[Newman(2004)]{newman_main}
M.~E.~J. Newman.
\newblock Coauthorship networks and patterns of scientific collaboration.
\newblock \emph{Proceedings of the National Academy of Sciences}, 101\penalty0
  (Supplement 1):\penalty0 5200--5205, April 2004.
\newblock ISSN 0027-8424, 1091-6490.
\newblock \doi{10.1073/pnas.0307545100}.
\newblock URL \url{http://www.pnas.org/cgi/doi/10.1073/pnas.0307545100}.

\bibitem[Wagner and Leydesdorff(2005)]{wagner_gener}
Caroline~S. Wagner and Loet Leydesdorff.
\newblock Network structure, self-organization, and the growth of international
  collaboration in science.
\newblock \emph{Research Policy}, 34\penalty0 (10):\penalty0 1608--1618,
  December 2005.
\newblock ISSN 00487333.
\newblock \doi{10.1016/j.respol.2005.08.002}.
\newblock URL
  \url{http://linkinghub.elsevier.com/retrieve/pii/S0048733305001745}.

\bibitem[Bassett and Bullmore(2006)]{swbrain}
Danielle~Smith Bassett and Ed~Bullmore.
\newblock Small-world brain networks.
\newblock \emph{The Neuroscientist}, 12\penalty0 (6):\penalty0 512--523, 2006.
\newblock \doi{10.1177/1073858406293182}.
\newblock URL \url{https://doi.org/10.1177/1073858406293182}.
\newblock PMID: 17079517.

\bibitem[Lužar et~al.(2014)Lužar, Levnajić, Povh, and Perc]{comscience}
Borut Lužar, Zoran Levnajić, Janez Povh, and Matjaž Perc.
\newblock Community {structure} and the {evolution} of {interdisciplinarity} in
  {Slovenia}'s {scientific} {collaboration} {network}.
\newblock \emph{PLoS ONE}, 9\penalty0 (4):\penalty0 e94429, April 2014.
\newblock ISSN 1932-6203.
\newblock \doi{10.1371/journal.pone.0094429}.
\newblock URL \url{http://dx.plos.org/10.1371/journal.pone.0094429}.

\bibitem[Gewin(2013)]{headstart}
Virginia Gewin.
\newblock Neuroscience: {A} head start for brain imaging.
\newblock \emph{Nature}, 503\penalty0 (7474):\penalty0 153--155, November 2013.
\newblock ISSN 0028-0836, 1476-4687.
\newblock \doi{10.1038/nj7474-153a}.
\newblock URL \url{http://www.nature.com/doifinder/10.1038/nj7474-153a}.

\bibitem[Rhoten(2004)]{interdisc}
D.~Rhoten.
\newblock {Risks} and {rewards} of an {interdisciplinary} {research} {path}.
\newblock \emph{Science}, 306\penalty0 (5704):\penalty0 2046--2046, December
  2004.
\newblock ISSN 0036-8075, 1095-9203.
\newblock \doi{10.1126/science.1103628}.
\newblock URL \url{http://www.sciencemag.org/cgi/doi/10.1126/science.1103628}.

\bibitem[Jacobs and Frickel(2009)]{interdisc_assess}
Jerry~A. Jacobs and Scott Frickel.
\newblock Interdisciplinarity: {A} {critical} {assessment}.
\newblock \emph{Annual Review of Sociology}, 35\penalty0 (1):\penalty0 43--65,
  August 2009.
\newblock ISSN 0360-0572, 1545-2115.
\newblock \doi{10.1146/annurev-soc-070308-115954}.
\newblock URL
  \url{http://www.annualreviews.org/doi/10.1146/annurev-soc-070308-115954}.

\bibitem[Davis(2007)]{davis_grand}
Lennard~J Davis.
\newblock A grand unified theory of interdisciplinarity.
\newblock \emph{Chronicles of Higher Education}, 53\penalty0 (40):\penalty0 B9,
  2007.

\bibitem[nas(2004)]{nasint}
\emph{Facilitating {interdisciplinary} {research}}.
\newblock National Academies Press, Washington, D.C., April 2004.
\newblock ISBN 978-0-309-09435-1.
\newblock URL \url{http://www.nap.edu/catalog/11153}.
\newblock DOI: 10.17226/11153.

\bibitem[Lopez(2011)]{alzburd}
Oscar~L. Lopez.
\newblock The growing burden of {Alzheimer}'s disease.
\newblock \emph{The American Journal of Managed Care}, 17 Suppl 13:\penalty0
  S339--345, November 2011.
\newblock ISSN 1936-2692.

\bibitem[Hughes et~al.(2014)Hughes, Peeler, Hogenesch, and
  Trojanowski]{alzcent}
Michael~E. Hughes, John Peeler, John~B. Hogenesch, and John~Q. Trojanowski.
\newblock The {Growth} and {Impact} of {Alzheimer} {Disease} {Centers} as
  {Measured} by {Social} {Network} {Analysis}.
\newblock \emph{JAMA Neurology}, 71\penalty0 (4):\penalty0 412, April 2014.
\newblock ISSN 2168-6149.
\newblock \doi{10.1001/jamaneurol.2013.6225}.
\newblock URL
  \url{http://archneur.jamanetwork.com/article.aspx?doi=10.1001/jamaneurol.2013.6225}.

\bibitem[Weiner et~al.(2015)Weiner, Veitch, Aisen, Beckett, Cairns, Cedarbaum,
  Donohue, Green, Harvey, Jack, Jagust, Morris, Petersen, Saykin, Shaw,
  Thompson, Toga, and Trojanowski]{adni}
Michael~W. Weiner, Dallas~P. Veitch, Paul~S. Aisen, Laurel~A. Beckett, Nigel~J.
  Cairns, Jesse Cedarbaum, Michael~C. Donohue, Robert~C. Green, Danielle
  Harvey, Clifford~R. Jack, William Jagust, John~C. Morris, Ronald~C. Petersen,
  Andrew~J. Saykin, Leslie Shaw, Paul~M. Thompson, Arthur~W. Toga, and John~Q.
  Trojanowski.
\newblock Impact of the {Alzheimer}'s {Disease} {Neuroimaging} {Initiative},
  2004 to 2014.
\newblock \emph{Alzheimer's \& Dementia}, 11\penalty0 (7):\penalty0 865--884,
  July 2015.
\newblock ISSN 15525260.
\newblock \doi{10.1016/j.jalz.2015.04.005}.
\newblock URL
  \url{http://linkinghub.elsevier.com/retrieve/pii/S1552526015001715}.

\end{thebibliography}

\clearpage

\appendix
\newcommand{\hbAppendixPrefix}{S}
\renewcommand{\thefigure}{\hbAppendixPrefix\arabic{figure}}
\setcounter{figure}{0}
\renewcommand{\thetable}{\hbAppendixPrefix\arabic{table}} 
\setcounter{table}{0}

\end{document}